\documentclass[prd,aps,preprint,showpacs]{revtex4}

\usepackage{axodraw}
\usepackage{epsfig}
\usepackage{color}

\newcommand{\cred}{\color{red}}

\def\bea{\begin{eqnarray}}
\def\eea{\end{eqnarray}}
\def\be{\begin{equation}}
\def\ee{\end{equation}}

\begin{document}

\begin{titlepage}
\voffset 1.5cm \preprint{ KIAS--P04041 \hspace{1ex}
}

\title{
Neutrino Mass from R-parity Violation in Split Supersymmetry}

\author{
Eung Jin Chun and Seong Chan Park}

\affiliation{ Korea Institute for Advanced Study, 207-43
Cheongryangri-dong, Dongdaemun-gu,  Seoul 130-722, Korea  }


\begin{abstract}
We investigate how the observed neutrino data can be accommodated
by R-parity violation in Split Supersymmetry. The atmospheric
neutrino mass and mixing are explained by the bilinear parameters
$\xi_i$ inducing the neutrino-neutralino mixing  as in the usual
low-energy supersymmetry.  Among various one-loop corrections,
only  the quark-squark exchanging diagrams involving the order-one
trilinear couplings $\lambda'_{i23,i32}$ can generate the solar
neutrino mass and mixing if the scalar mass $m_S$ is not larger
than $10^9$ GeV. This scheme requires an unpleasant hierarchical
structure of the couplings, e.g., $\lambda_{i23,i32}\sim 1$,
$\lambda'_{i33} \lesssim 10^{-4}$ and $\xi_i \lesssim 10^{-6}$. On
the other hand, the model has a distinct collider signature of the
lightest neutralino which can decay only to the final states, $l_i
W^{(*)}$ and $\nu Z^{(*)}$, arising from the bilinear mixing.
Thus, the measurement of the ratio; $\Gamma(e W^{(*)}) :
\Gamma(\mu W^{(*)}) : \Gamma(\tau W^{(*)})$ would provide a clean
probe of the small reactor and large atmospheric neutrino mixing
angles as far as the neutralino mass is larger than 62 GeV.
\end{abstract}

\maketitle
\end{titlepage}

The gauge hierarchy problem has been considered as a strong
motivation for low-energy supersymmetry, in which all the
observable sector soft parameters lie in the TeV scale, ensuring
the naturalness of the Higgs mass.  However, a rather radical
suggestion has been made  to abandon the naturalness property and
consider a high-scale supersymmetry \cite{AD}, in which all the
scalars are extremely heavy, except for the finely tuned Higgs
bosons, and the fermions (including gauginos and Higgsinos) remain
light. This approach, dubbed as ``Split Supersymmetry'' \cite{GR},
was further advocated by gauge unification and dark matter taken
as the guiding principle for new physics. The idea of Split
Supersymmetry, being trivially free from the difficulties like
flavor-changing neutral current and CP problems and the
cosmological gravitino problem, predicts a distinct phenomenology
of a long-lived gluino, which could be probed in the future
collider or cosmological experiments \cite{Glu}.

In this paper, we wish to investigate the possibility of allowing
R-parity violation as the origin of the neutrino masses and mixing
\cite{HS} in the framework of Split Supersymmetry.  Actually the
heaviness of the squarks and sleptons allows the large R-parity
violation without any difficulty with low energy precision
measurements and this interesting feature of Split Supersymmetry
is crucially used to produce the proper neutrino masses and
mixing. Having abandoned R-parity conservation, the lightest
supersymmetric particle is now destabilized, and thus one may have
to abandon a nice dark matter candidate, a neutralino. Indeed, the
neutralino decays very fast if R-parity violation accounts for the
observed neutrino masses not only in the conventional low-energy
supersymmetry but also in Split Supersymmetry under discussion
\cite{GKM}.  Even though we loose the dark matter as the guiding
principle \cite{GR}, the instability of the lightest
supersymmetric particle in the TeV scale neutralino sector could
provide another way of probing the idea of Split Supersymmetry
combined with the origin of neutrino masses and mixing, as is well
known in the conventional framework \cite{MRV,jaja,PHRV}.  As we
will see, there are several different features in generating the
neutrino mass matrix from R-parity violation in Split
Supersymmetry summarized as follows.
\begin{itemize}
\item
Tree-level neutrino mass matrix pattern, coming from the mixing
between neutrinos and neutralinos, is the same as in low-energy
supersymmetry.  This requires small bilinear parameters; $\xi_i
\lesssim 10^{-6}$.
\item
The effect of the slepton--Higgs mixing is negligible as the
scalar masses are very high. As a result,  it is not possible to
explain observed neutrino data only by bilinear terms.
\item
The solar neutrino mass and mixing can only be generated by the
usual one-loop diagrams involving down-type quarks and squarks
with the trilinear couplings $\lambda'_{i23}$ and $\lambda'_{i32}$
of the order one.

\item
The lightest neutralino can decay only to the gauge bosons,
$W^{\pm(*)}$ and $Z^{0(*)}$, through which the tree-level neutrino
mass parameters could be probed in the collider experiments for
the neutralino mass larger than about 62 GeV.

\item
The model requires an ad-hoc hierarchical structure of the
couplings, namely, $\lambda'_{i23, i32} \sim 1$, $\lambda'_{i33}
\lesssim 10^{-4}$ and $\lambda_{i33} \lesssim 10^{-3}$, where the
last two constraints come from the limit of the bilinear
parameter, $\xi_i \lesssim 10^{-6}$.
\end{itemize}
The last point should be contrasted with the case of the
low-energy supersymmetry models, where the neutrino data can well
be explained by assuming the usual hierarchy of the trilinear
couplings $\lambda_{ijk}, \lambda'_{ijk} \leq \lambda'_{i33},
\lambda_{i33} \sim 10^{-4} - 10^{-5}$, which could be the
consequence of a family $U(1)$ symmetry \cite{choi}.

\medskip

Let us now start our main discussion by considering first the
features of the bilinear R-parity violation in Split
Supersymmetry. The gauge invariant bilinear terms in the
superpotential and the scalar potential are
\begin{eqnarray} \label{BiTerms}
 W &=& \mu H_1 H_2 + \epsilon_i \mu L_i H_2 \,, \nonumber\\
 V &=& B H_1 H_2 +  B_i L_i H_2
 + m^2_{L_iH_1} L_i H_1^\dagger +h.c.\,,
\end{eqnarray}
where we have used the same notation, $H_{1,2}$ and $L_i$, for the
superfields and their scalar components of the Higgs and lepton
doublets.  In Split Supersymmetry, the  dimension-two soft
parameters $B$ or  $B_i$ and $m^2_{L_iH_1}$ could  be of the order
$m_S^2$ or $\epsilon_i m_S^2$ where the high-scale of the scalar
masses is likely to be in the range: $m_S = 10^9 - 10^{13}$ GeV
\cite{AD,GR}.

From the above potential (\ref{BiTerms}), one finds the following
R-parity violating parameter;
\begin{equation} \label{xiis}
\xi_i \equiv {\langle \tilde{\nu}_i \rangle \over \langle H_1^0
\rangle } - \epsilon_i = {m^2_{L_iH_1} \over m^2_{L_i} } + {B_i
\over m^2_{L_i}} t_\beta - \epsilon_i\,,
\end{equation}
where $t_\beta = \tan\beta = \langle H_2^0 \rangle/ \langle H_1^0
\rangle$ and $m_{L_i}\sim m_S$ is the soft mass of the $i$-th
slepton. The above parameter $\xi_i$ determines the well-known
mixing between the leptons and the gauginos/Higgsinos giving rise
to the tree-level neutrino mass matrix (see Fig. ~1);
\begin{equation} \label{M0}
M^{\nu\, (0)}_{ij} = -{M_Z^2 \over F_N } \xi_i \xi_j c_\beta^2
\end{equation}
where $F_N \equiv M_1 M_2/(c_W^2 M_1 + s_W^2 M_2) + M_Z^2
s_{2\beta} / \mu$ is the mass parameter deduced from the 4x4
neutralino mass matrix \cite{jaja}.  For the atmospheric neutrino
mass scale, $m_{\nu_3} = \sqrt{\Delta m^2_{atm}}  \sim 0.05$ eV,
one determines the size of $\xi\equiv \sqrt{\sum_i |\xi_i|^2}$ :
\begin{equation} \label{cond1}
\xi c_\beta = 7.4\times10^{-7} \left( F_N \over M_Z
\right)^{1\over2} \left( m_{\nu_3} \over 0.05 \mbox{ eV}
\right)^{1\over2} .
\end{equation}
Barring the cancellation among the three terms on the right-hand
side of Eq.~(\ref{xiis}), this implies that each term  should not
be larger than  $10^{-6}/c_\beta$.

\begin{figure}
\begin{center}
\begin{minipage}{5cm}
\begin{picture}(300,56)(100,0)
    \Vertex(120,10){1.5}
    \Vertex(180,10){1.5}
    \put(100,15){$\cred \nu_i$}
    \put(195,15){$\cred \nu_j$}
    \put(115,0){$\cred \xi_i$}
    \put(175,0){$\cred \xi_j$}
    \SetColor{Brown}
    \ArrowLine(100,10)(120,10) 
    \ArrowLine(200,10)(180,10)  
    \ArrowLine(150,10)(120,10)
    \ArrowLine(150,10)(180,10)
    \put(150,20){$\chi^0$}
    \SetColor{Brown}
    \SetColor{Black}
    \Line(147,7)(153,13)
    \Line(147,13)(153,7)
\end{picture}
\end{minipage}
\end{center}
\caption{ Tree-level diagrams generating neutrino masses which are
induced by the mixing between the neutrinos and
gauginos/higgsinos.}
\end{figure}
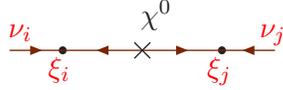

\medskip

When the mass matrix (\ref{M0}) explains the atmospheric neutrino
data, the atmospheric neutrino mixing angle $\theta_{23}$ and the
reactor neutrino mixing angle $\theta_{13}$ are approximately
determined as follows \cite{jaja}:
\begin{eqnarray} \label{thetas}
\sin^22\theta_{23} &\approx& 4 \, |\hat{\xi}_2|^2 \,
|\hat{\xi}_3|^2
\nonumber\\
\sin^22\theta_{13} &\approx& 4 \, |\hat{\xi}_1|^2\left( 1-
|\hat{\xi}_1|^2\right)
\end{eqnarray}
where $\hat{\xi}_i$ is the unit vector of $\xi_i$: $\hat{\xi}_i
\equiv \xi_i/\xi$. This feature can be probed in collider
experiments as will be discussed later.

While the tree mass matrix (\ref{M0}) can generate only one
nonzero mass eigenvalue, presumably $m_{\nu_3}$, the second
largest mass, $m_{\nu_2} = \sqrt{\Delta m^2_{sol}} \sim 9$ meV,
can be generated from one-loop radiative corrections. Combining
these, both the atmospheric and the solar neutrino data can be
accommodated in the conventional supersymmetric theories
\cite{bimodel,bibi}.  There is another type of bilinear parameters
defined by
\begin{equation} \label{etais}
\eta_i \equiv \xi_i - {B_i \over B},
\end{equation}
which quantifies  the mixing between the sleptons and Higgs bosons
and appears in the one-loop diagrams.  The parameter $\eta_i$ can
play an important role to produce sizable one-loop contribution in
the usual supersymmetric theories \cite{bibi}. However, this is
not the case in Split Supersymmetry  where the one-loop
contributions always come with the combination of
\begin{equation} \label{condeta}
 \eta_i \, {m_Z^2\over m_{L_i}^2}
\end{equation}
which becomes very small for  $m_Z \ll m_S$. Therefore, the
bilinear parameter $\xi_i$ alone cannot explain the observed
neutrino data, which excludes the bilinear model as the origin of
the neutrino masses and mixing in the context of Split
Supersymmetry.

\medskip

Let us ask whether the inclusion of trilinear R-parity violating
can be a viable option for the generation of the desired neutrino
mass matrix. The general lepton number violating trilinear terms
in the  superpotential are
\begin{equation} \label{Wtri}
W= \lambda'_{ijk} L_i Q_j D^c_k + \lambda_{ijk} L_i L_j E^c_k
\end{equation}
where we take $i<j$ for $\lambda_{ijk}$ which is antisymmetric
under the exchange, $i\leftrightarrow j$.  For our discussion, it
is important to realize that certain trilinear couplings can
radiatively generate the bilinear parameters (2), and thus can be
strongly constrained.  The most strongly constrained couplings are
$\lambda'_{i33}$ or $\lambda_{i33}$ which contribute to the
renormalization group evolution  of the bilinear parameter, e.g.,
$\epsilon_i$ as follows \cite{dreiner}:
\begin{equation}
16 \pi^2 {d \epsilon_i \over d t} \sim -3 \lambda'_{ijj} h_{d_j} -
\lambda_{ijj} h_{e_j}\,.
\end{equation}
Solving the above equation by one-step approximation, we find
\begin{equation}
\epsilon_i c_\beta = {1\over 16\pi^2}(3 \lambda'_{ijj}
{m_{d_j}\over v } + \lambda_{ijj} {m_{e_j} \over v} ) \ln{M_X
\over m_S}.
\end{equation}
For the values of $M_X=10^{16}$ GeV and $m_S=10^{9}$ GeV, and the
condition for the bilinear parameter $|\xi_i|c_\beta \approx
|\epsilon_i|c_\beta$ (\ref{cond1}), we obtain
\begin{equation} \label{lambound}
\lambda'_{i11}\, {m_d \over m_b},\;  \lambda'_{i22}\, {m_s \over
m_b},\; \lambda'_{i33} \lesssim 10^{-4} \,.
\end{equation}
for $F_N=M_Z$.
 Applying the similar argument, the bounds on the
couplings $\lambda_{ijj}$ are found to be
\begin{equation} \label{lambound3}
\lambda_{i11}\, {m_e \over m_\tau},\;  \lambda_{i22}\, {m_\mu
\over m_\tau},\; \lambda_{i33} \lesssim 7 \times 10^{-4} \,.
\end{equation}
Such small trilinear couplings (\ref{lambound},\ref{lambound3})
cannot generate a sizable one-loop correction to the neutrino
masses explaining the solar neutrino data.  As an example,
consider the typical one-loop contribution to neutrino masses
coming from the quark-squark exchange (see Fig. ~2):
\begin{equation} \label{M1b}
 M^{\nu\, (1)}_{ij} \approx {3\over 8 \pi^2}
 \lambda'_{ikl}\lambda'_{jlk} { m_{d_k} m_{d_l} A_d
 \over m_S^2}
\end{equation}
where we have taken the squark mass to be $m_S$.

\begin{figure}
\begin{center}
\begin{minipage}{5cm}
\begin{picture}(300,56)(100,0)
    \Vertex(120,10){1.5}
    \Vertex(180,10){1.5}
    \put(100,15){$\cred \nu_i$}
    \put(195,15){$\cred \nu_j$}
    \put(115,0){$\cred \lambda'_{ikl}$}
    \put(175,0){$\cred \lambda'_{jlk}$}
    \SetColor{Brown}
    \ArrowLine(100,10)(120,10) 
    \ArrowLine(200,10)(180,10)  
    \ArrowLine(150,10)(120,10)
    \ArrowLine(150,10)(180,10)
    \put(135,13){$d^c_l$}
    \put(160,13){$d_l$}
    \SetColor{Brown}
    \DashArrowArcn(150,10)(30,90,0){3}
    \DashArrowArc(150,10)(30,90,180){3}
    \put(118,33){$\tilde{d}_k$}
    \put(175,33){$\tilde{d^c}_k$}
    \SetColor{Black}
    \Line(147,7)(153,13)
    \Line(147,13)(153,7)
    \Line(147,37)(153,43)
    \Line(147,43)(153,37)
\end{picture}
\end{minipage}
\end{center}
\caption{One-loop diagrams generating sizable neutrino masses from
the quark-squark exchange. }
\end{figure}
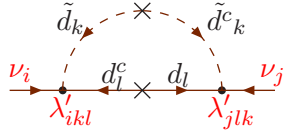

\medskip

Taking $\lambda'_{i33} =10^{-4}$ and $A=m_S=10^9$ GeV, one finds
that the one-loop mass  (\ref{M1b}) become far below the solar
neutrino mass scale $\sim 9$ meV.    The only way to obtain such a
sizable neutrino mass scale is to have order-one trilinear
couplings avoiding the above constraint (\ref{lambound}).  It
turns out that the unique possibility for this is to allow the
couplings
$$ \lambda'_{i23} \quad\mbox{and}\quad \lambda'_{i32} \,.$$
The above couplings can induce the bilinear parameter at two-loop
level, which can be viewed as the generation of the induced
coupling of $\lambda'_{i33}$ \cite{dreiner}:
\begin{equation}
 \lambda^{\prime}_{i33}\big|_{induced} \sim {1\over 16\pi^2} \lambda'_{i32}
 h_t^2 V_{ts} {m_s \over m_b} \quad\mbox{or}\quad
 {1\over 16\pi^2} \lambda'_{i23}
 h_t^2 V_{ts}
\end{equation}
where $V_{ts}$ is the CKM mixing element of quarks.  Thus, the
couplings $ \lambda'_{i23}$ and $\lambda'_{i32}$ can be of the
order one and can generate the following neutrino mass elements:
\begin{equation} \label{M1bs}
 M^{\nu\, (1)}_{ij} \approx {3\over 8 \pi^2}
 \lambda'_{i23}\lambda'_{j32} { m_b m_s  \over m_S} \sim
 10^{-2}\;\mbox{eV}
\end{equation}
for $m_S\sim 10^9$ GeV.  For much larger $m_S$, the above
contribution becomes much too small and cannot  be used for our
purpose. The elements $M^{\nu\,(1)}_{ij}$ with ($i,j=1,2$) in Eq.
(\ref{M1bs}) essentially determine the large solar neutrino mixing
angle $\theta_{12}$ which can be arranged appropriate by the
choice of the trilinear couplings;
$\lambda'_{i23}\lambda'_{j32}\sim 1$ for all $i$ and $j$. We do
not bother to present the detailed values for this as it will not
be necessary in the following discussions.

\medskip

Concerning the $\lambda$ couplings, the leading contributions come
from the diagrams in Fig.~3 which do not have the $1/m_S$
suppression  \cite{kang}.  Given the constraints
(\ref{lambound3}), only the couplings $\lambda_{ikl}$ (with $i\neq
k\neq l$) can be large enough to generate sizable neutrino mass
components. Calculation of the diagrams in Fig.~3 gives us
\begin{equation} \label{M1l}
M^{\nu\,(1)}_{ik}  \approx {g \over 8\pi^2} \lambda_{ikl} \xi_l
c_\beta^2  {m_{e_l}  M_W \over \sqrt{2} \mu} \, \ln{m_{L_i} \over
m_{L_k}}
\end{equation}
which is a fairly good approximation in the typical region of
parameter space: $M_2,\mu > M_W$ and $t_\beta>3$.  Comparing this
with  the tree contributions in Eq.~(\ref{M0}), we  get the
loop-to-tree ratio:
\begin{equation} \label{mratio}
 {m_{\nu_2} \over m_{\nu_3}} \approx {g\over 8 \pi^2} {m_{e_l} M_W \over
 M_Z^2} { \lambda_{ikl} \xi_l \over |\xi|^2 } { F_N \over \mu}\,,
\end{equation}
which shows that  the neutrino mass elements of the order of the
solar neutrino mass scale $m_{\nu_2} \sim 9$ meV can be obtained
if the couplings are as large as
\begin{equation} \label{cond2}
\lambda_{ikl}{m_{e_l}\over m_\tau} \sim 10^3 \xi \sim 10^{-3}
t_\beta \,.
\end{equation}
From this, one finds that $\lambda_{231}$ has no effect due to the
strong suppression by $m_e/m_\tau$ and thus only sizable $M_{12}$
and $M_{13}$ components  can be generated  from $\lambda_{123}$
and $\lambda_{132}$, respectively.  However, this pattern cannot
accommodate the solar neutrino mass-squared difference and the
mixing angle properly.

\medskip

Therefore, we conclude that the atmospheric and solar neutrino
data can be explained by the combination of the tree-level and the
loop contribution given by in Eqs~(\ref{M0}) and (\ref{M1bs}),
respectively, in the context of Split Supersymmetry.  For this, we
need the small bilinear couplings and order-one trilinear
couplings:
\begin{eqnarray}
\xi_1 \ll \xi_2 \approx \xi_3 \sim 10^{-6} \quad\mbox{and}\quad
\lambda'_{i32},\lambda'_{i23} \sim 1\,,
\end{eqnarray}
and the scalar masses of the order $m_S \sim 10^9$ GeV. Note that
the bilinear parameter $\xi$ can come from the tree-level input
$\epsilon_i$ of the same order, or from the radiative generation
by the trilinear coupling $\lambda'_{i33}$ of the order $10^{-4}$
as in Eq.~(\ref{lambound}).

\begin{figure}
\begin{center}
\begin{minipage}{5cm}
\begin{picture}(300,56)(100,0)
    \Vertex(120,10){1.5}
    \Vertex(180,10){1.5}
    \put(100,15){$\cred \nu_i$}
    \put(195,15){$\cred \nu_k$}
    \put(115,0){$\cred \lambda_{ikl}$}
    \put(175,0){$\cred g$}
    \SetColor{Brown}
    \ArrowLine(100,10)(120,10) 
    \ArrowLine(200,10)(180,10)  
    \ArrowLine(150,10)(120,10)
    \ArrowLine(150,10)(180,10)
    \put(135,13){$e^c_l$}
    \put(160,13){$\tilde{W}^-$}
    \SetColor{Brown}
    \DashArrowArc(150,10)(30,0,90){3}
    \DashArrowArc(150,10)(30,90,180){3}
    \put(118,33){$\tilde{e}_k$}
    \put(175,33){$\tilde{e}_k$}
    \SetColor{Black}
    \Line(147,7)(153,13)
    \Line(147,13)(153,7)
\end{picture}
\end{minipage}
\begin{minipage}{5cm}
\begin{picture}(300,56)(100,0)
    \Vertex(120,10){1.5}
    \Vertex(180,10){1.5}
    \put(100,15){$\cred \nu_k$}
    \put(195,15){$\cred \nu_i$}
    \put(115,0){$\cred -\lambda_{ikl}$}
    \put(175,0){$\cred g$}
    \SetColor{Brown}
    \ArrowLine(100,10)(120,10) 
    \ArrowLine(200,10)(180,10)  
    \ArrowLine(150,10)(120,10)
    \ArrowLine(150,10)(180,10)
    \put(135,13){$e^c_l$}
    \put(160,13){$\tilde{W}^-$}
    \SetColor{Brown}
    \DashArrowArc(150,10)(30,0,90){3}
    \DashArrowArc(150,10)(30,90,180){3}
    \put(118,33){$\tilde{e}_i$}
    \put(175,33){$\tilde{e}_i$}
    \SetColor{Black}
    \Line(147,7)(153,13)
    \Line(147,13)(153,7)
\end{picture}
\end{minipage}
\end{center}
\caption{One-loop diagrams generating sizable neutrino masses with
no ultra heavy mass suppression but with lepton mass suppression.
}
\end{figure}
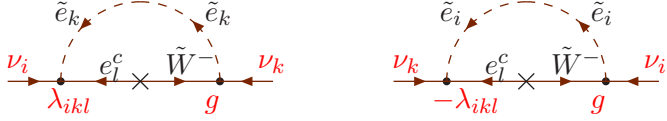

\medskip

Split Supersymmetry predicts distinct signatures from the R-parity
violating decays of the lightest neutralino, $\chi_1^0$, which are
different from the conventional supersymmetric theories. Similarly
to the gluino case, the neutralino decay processes through squark
exchange are highly suppressed.  But, the lightest neutralino
decay can occur through the bilinear mixing of leptons and
neutralinos governed by the parameter $\xi_i$.  This  has to be
contrasted with the low-energy supersymmetry case where the effect
of trilinear couplings can be sizable to be detected in the future
collider experiments \cite{Jung}.  Let us discuss if the
neutralino decay length is short enough to allow the observation
of its decay modes in the colliders.   When $\chi_1^0$ is heavier
than $Z$, it decays to $lW$ and $\nu Z$ having the following
rates:
\begin{eqnarray} \label{2body}
 \Gamma(l_i W) &\approx& {G_F m_{\chi_1^0}^3 \over 4\sqrt{2} \pi }
 |\xi_i|^2 c_\beta^2 {\cal I}(\frac{m_W^2}{m_{\chi_1^0}^2}), \nonumber \\
 \Gamma(\nu_i Z)&\approx& {G_F m_{\chi_1^0}^3 \over 16\sqrt{2} \pi }
 |\xi_i|^2 c_\beta^2 {\cal I}(\frac{m_Z^2}{m_{\chi_1^0}^2}), \\
 \Gamma_{2-body} = \Gamma(l W + \nu Z) & \approx &
  {G_F m_{\chi_1^0}^3 \over 16 \sqrt{2} \pi}
 |\xi|^2 c_\beta^2 \left(4 {\cal I}(\frac{m_W^2}{m_{\chi_1^0}^2})
 +{\cal I}(\frac{m_Z^2}{m_{\chi_1^0}^2})\right) \,. \nonumber
\end{eqnarray}
where we have ignored the fermion masses and the kinematical
factors are given as follows.
\begin{eqnarray}
{\cal I}(x)\equiv \left(1- x\right)^2\left(1+2
x\right)\theta(x-1). \nonumber
\end{eqnarray}
When $\chi_1^0$ is lighter than $W$, it has only 3-body decay
modes through the virtual $W$ and $Z$ and its decay rates are then
\begin{eqnarray} \label{3body}
\Gamma(l_i W^*) &\approx& {3G_F^2 m_{\chi_1^0}^5 \over 64 \pi^3 }
 |\xi_i|^2 c_\beta^2 \\
 \Gamma_{3-body}=\Gamma(lW^*+\nu Z^*) &\approx &
  {5.4 G_F^2 m_{\chi_1^0}^5 \over 64 \pi^3 }
 |\xi|^2 c_\beta^2 \,. \nonumber
\end{eqnarray}
From the above equations (\ref{2body}) and (\ref{3body}),  one
finds the decay length;
$$
\frac{1}{\Gamma}= 1.8 \, mm \quad \mbox{or} \quad  1\, m \,.
$$
for $|\xi|c_\beta=7.4\times10^{-7}$ (\ref{cond1}) and
$m_{\chi_1^0} = 100$  GeV or 62 GeV, respectively. Thus, the
neutralino mass is required to be larger than 62 GeV.
Independently of the 2-body or 3-body decay case, the measurement
of the branching ratios for the mode $l_i W$ will determine the
relative sizes of $|\xi_i|$
$$
|\xi_1|^2 : |\xi_2|^2 : |\xi_3|^2 = \mbox{B}(e W^{(*)}) :
\mbox{B}(\mu W^{(*)}) : \mbox{B}(\tau W^{(*)})
$$
which could provide the test of the model predicting the relation
(\ref{thetas}).  In the usual low-energy supersymmetry, the above
conclusion may not be secured if the neutralino allows only the
3-body decay modes for which the effect of trilinear couplings can
be even larger than the bilinear effect \cite{Jung}.

\medskip

In conclusion, we have shown how the observed neutrino data can be
accommodated in the context of Split Supersymmetry with R-parity
violation.  As most one-loop diagrams generating the neutrino mass
matrix are suppressed by the high scale of the scalar masses,
there appears a rather unique way to accommodate the desired
neutrino mass matrix.  The tree-level neutrino mass matrix coming
from the bilinear parameters can nicely explain the atmospheric
neutrino mass and mixing by the same way as in the usual
low-energy supersymmetry. This requires the bilinear parameters to
be of the order $10^{-6}$ and put some bounds on the trilinear
couplings like $\lambda'_{i33} \lesssim 10^{-4}$. Among various
one-loop corrections, the quark-squark exchange diagrams can
produce the solar neutrino mass and mixing taking the order-one
couplings $\lambda'_{i23,i32}$ if the scalar mass is at its lower
end:  $m_S \sim 10^9$ GeV.

In this scheme, the lightest neutralino can decay only through the
bilinear mixing into the final states of $l_i W^{(*)}$ and $\nu
Z^{(*)}$.  The corresponding decay length can be less than 1 m if
the neutralino mass is larger than 62 GeV.  Thus, the observation
of such features and the measurement of the branching ratios
following the relation: $B(e W^{(*)}) \ll B(\mu W^{(*)}) \approx
B(\tau W^{(*)})$ would  provide a clean probe of the model
prediction coming from the small reactor and large atmospheric
neutrino mixing angles.

We would add some words on fine tuning in the parameter space to
explain neutrino masses.  The seemingly unnatural parameter space
with the small bilinear couplings ($\lesssim 10^{-6}$ and the
hierarchy in the trilinear couplings (e.g., $\lambda'_{i33} \ll
\lambda'_{i23,i32}\sim 1 $) is rather uniquely chosen to fit the
experimental data of neutrino masses and mixing. This makes more
complicate the Yukawa hierarchy problem, which appears to be the
generic feature of the R-parity violating Split Supersymmetry.

\medskip

{\bf Acknowledgement}:  EJC was supported by the Korea Research
Foundation Grant, KRF-2002-070-C00022. SC thanks to S.Y.Choi for
useful discussions.


\end{document}